\documentclass[preprintnumbers,superscriptaddress,showpacs,pra]{revtex4}
\usepackage{amsmath}
\usepackage{epsfig}
\usepackage{graphicx}
\usepackage{color}

\input{tcilatex}
\begin{document}

\title[Geometric Potential from Quantum Conditions]{Identification of
Geometric Potential from Quantum Conditions for a Particle on a Curved
Surface}
\author{D. K. Lian, L. D. Hu, and Q. H. Liu}
\affiliation{School for Theoretical Physics, School of Physics and Electronics, Hunan
University, Changsha 410082, China}
\date{\today }

\begin{abstract}
Combination of a construction of unambiguous quantum conditions out of the
conventional one and a simultaneous quantization of the positions, momenta,
angular momenta and Hamiltonian leads to the geometric potential given by
the so-called thin-lay quantization.
\end{abstract}

\pacs{03.65.Ca Formalism; 04.60.Ds Canonical
quantization;
02.40.-k
Geometry, differential geometry, and topology;
68.65.-k
Low-dimensional,
mesoscopic, and nanoscale systems: structure
and
nonelectronic properties; }
\maketitle

\section{Introduction}

In quantum mechanics for a system, the construction of a proper quantum
Hamiltonian operator takes the central position. For a free particle
constrained to live on a curved surface or a curved space, DeWitt in 1957
used a specific generalization of Feynman's time-sliced formula in Cartesian
coordinates and found a surprising result that his amplitude turned out to
satisfy a Schr\"{o}dinger equation different from what had previously
assumed by Schr\"{o}dinger \cite{Sch} and Podolsky \cite{podolsky}. In
addition to the kinetic term which is Laplace-Beltrami operator divided by
two times of mass, his Hamilton operator contained an extra effective
potential proportional to the intrinsic curvature scalar.

Jensen and Koppe\ in 1972 \cite{jk} and subsequently da Costa \cite{dacosta}
in 1981 developed a thin-layer quantization (also known as confining
potential formalism) to deal with the free motion on the curved surface and
demonstrated that the particle experiences a quantum potential that is a
function of the intrinsic and extrinsic curvatures of the curved surface,
which was later called the \textit{geometric potential }\cite{2004}. By the
thin-layer quantization we mean a treatment of ($n-1$)-dimensional smooth
surface $S^{n-1}$ in flat space $R^{n}$ ($n\succeq 1$) and two infinitely
high potential walls at the distance $\delta \rightarrow 0$ from the
surface. Since the excitation energies of the particle in the direction
normal to the surface are much larger than those in the tangential direction
so that the degree of freedom along the normal direction is actually frozen
to the ground state, an effective dynamics for the constrained system on the
surface is thus established. This thin-layer quantization has a distinct
feature for no presence of operator-ordering difficulty or other
ambiguities. It is thus a powerful tool to examine various curvature-induced
consequences in low-dimensional curved nanostructures, for instance,
spin-orbit interaction of electrons on a curved surface \cite{2001}, the
mechanical-quantum-bit states \cite{2004}, the geometry-induced charge
separation on helicoidal ribbon \cite{20091}, the curvature-induced p-n
junctions in bilayer graphene \cite{20092}, the periodic curvature dependent
electrical resistivity of corrugated semiconductor films \cite{20093} as
well as the geometry-driven shift in the Tomonaga-Luttinger liquid \cite%
{20094}, electronic band-gap opening in corrugated graphene \cite{2010},
low-temperature resistivity anomalies in periodic curved surfaces \cite%
{20101}, curvature effects in thin magnetic shells \cite{2014}, and the
induced magnetic moment for a spinless charged particle on a curved wire 
\cite{2015}, etc. \cite{packet1,packet2,packet3,packet4,packet5}
Experimental confirmations include: an optical realization of the geometric
potential \cite{exp1} in 2010 and the geometric potential in a
one-dimensional metallic $\mathit{C}_{60}$ polymer with an uneven periodic
peanut-shaped structure in 2012 \cite{exp2}. Applying the thin-layer
quantization to momentum operators which are fundamentally defined as
generators of a space translation, we have geometric momenta \cite{liu13-2}
which depends on the extrinsic curvatures of the curved surface.

It is generally accepted that the canonical quantization offers a
fundamental framework to directly construct the quantum operators, and the 
\textit{fundamental quantum conditions} are commutators\ between components
of position and momentum \cite{dirac1,dirac2}. Many explorations have been
devoted to searching for the \textit{geometric potential }within the
framework \cite%
{geoquan,homma,ogawa1,ogawa2,ikegami,mbbj,ILNC,kleinert,Tagirov,hong,Golovnev,2011,weinberg}%
. It is curious that no attempt is successful for even simplest
two-dimensional curved surface $S^{2}$ embedded in $R^{3}$. The best result
was to start from surface equation $df(x)/dt=0$, the time $t$ derivative of
the direct one $f(x)=0$ rather than $f(x)=0$ itself, to obtain a potential
depending on the the intrinsic and extrinsic curvatures via two arbitrary
real coefficients \cite{ikegami}. Some results are contradictory with each
other, for instance, for a free particle on a ($n-1$) dimensional sphere
Kleinert and Shabanov predicted no existence of any quantum potential \cite%
{kleinert}, but Hong and Rothe gave a quantum potential whose $n$-dependent
multiple is $(n+1)(n-3)$ \cite{hong}, whereas the thin-layer quantization
presented $(n-1)(n-3)$ for such a multiple \cite{ogawa1,ikegami,liu13-1}. We
revisited all these attempts, and concluded that the canonical quantization
together with Schr\"{o}dinger-Podolsky-DeWitt approach of Hamiltonian
operator construction was dubious, for the kinetic energy in it takes some
presumed forms that are primarily a sum of the Cartesian momenta squared.
Since 2011, we have tried to enlarge the canonical quantization scheme to
simultaneously quantize the Hamiltonian together with positions and momenta 
\cite{liu11}, rather than substituted the position and momentum operators
into some presumed forms of Hamiltonian. Yet the success is limited. i) We
obtained the geometric momentum which is identical to that given by the
thin-layer quantization \cite{liu13-2}, and ii) we got the correct form of
geometric potential for the ($n-1$) dimensional sphere \cite{liu13-1}, but
iii) there are ambiguities associated with geometric potential for other
curved surfaces \cite{liu17}.

In this Letter, we report that for a particle on a two-dimensional curved
surface, simultaneous quantization of the Hamiltonian together with the
fundamental quantities as position, momentum and the \textit{angular momentum%
}, the Hamiltonian includes correct form of the geometric potential. It is
the first time to achieve this result within the canonical quantization
scheme.

\section{Dirac's theory of the constrained systems: Classical and quantum
mechanics}

Let us consider a non-relativistically free particle that is constrained to
remain on a surface described by a constraint in configurational space $%
f(x)=0$, where $f(x)$ is some smooth function of position $x$, whose normal
vector is $\mathbf{n}\equiv \nabla f(x)/|\nabla f(x)|$. We can always choose
the equation of the surface such that $|\nabla f(x)|=1$, so that $\mathbf{n}%
\equiv \nabla f(x)$. In classical mechanics, the Hamiltonian is simply $%
H=p^{2}/2\mu $ where $p$ denotes the momentum, and $\mu $ denotes the mass.
However, in quantum mechanics, we can not impose the usual canonical
commutation relations $[x_{i},p_{j}]=i\hbar \delta _{ij}$, ($i,j=1,2,3$).
Dirac was aware of the fact the presence of this constraint which needed to
be eliminated before quantization could very well cause the remaining
classical phase to not admit Cartesian coordinates.

\subsection{Dirac brackets formulation of the classical motion}

Dirac gave a general theory for a large class of constrained Hamiltonian
systems including the motion on the surface \cite{dirac1}. He introduced a
bracket instead of the Poisson one $[f(x,p),g(x,p)]_{P}$ between any pair of
quantitates $f(x,p)$ and $g(x,p)$ in the following, 
\begin{equation}
\lbrack f(x,p),g(x,p)]_{D}\equiv \lbrack f(x,p),g(x,p)]_{P}-[f(x,p),\chi
_{\alpha }(x,p)]_{P}C_{\alpha \beta }^{-1}[\chi _{\beta }(x,p),g(x,p)]_{P},
\label{diracbracket}
\end{equation}%
where repeated indices are summed over in whole of this Letter, and $%
C_{\alpha \beta }\equiv \lbrack \chi _{\alpha }(x,p),\chi _{\beta }(x,p)]_{P}
$ are the matrix elements in the constraint matrix $\left( C_{\alpha \beta
}\right) $ and the functions $\chi _{\alpha }(q,p)$ ($\alpha ,\beta =1,2$)
are two constraints \cite{weinberg}, 
\begin{equation}
\chi _{1}(x,p)\equiv f(x)\left( =0\right) ,\text{and }\chi _{2}(x,p)\equiv 
\mathbf{n\cdot p}\left( =0\right) \mathbf{.}  \label{constraints}
\end{equation}%
The bracket $[f(x,p),g(x,p)]_{D}$ is called the Dirac bracket. We have
elementary Dirac brackets in the following, with use of symbol $%
n_{i,j}\equiv \partial n_{i}/\partial x_{j}$ \cite{weinberg},%
\begin{align}
\lbrack x_{i},x_{j}]_{D}& =0,  \label{xx} \\
\lbrack x_{i},p_{j}]_{D}& =\delta _{ij}-n_{i}n_{j},  \label{xp} \\
\lbrack p_{i},p_{j}]_{D}& =(n_{j}n_{i,k}-n_{i}n_{j,k})p_{k}.  \label{pp}
\end{align}%
These brackets were in general taken the fundamental set which after
quantization forms the set of the so-called fundamental quantum conditions 
\cite{dirac1,dirac2}. In classical mechanizes, the motion particle follows
the geodesic whose curvature is $\kappa $ and torsion is $\tau $. Now we
define the orbital angular momentum $\mathbf{G}\equiv \mathbf{x}\times 
\mathbf{p}$ for a particle on a curved surface. We do not use the
conventional symbol $\mathbf{L}$ that is usually used to denote the orbital
angular momentum where three components form a so(3) algebra $%
[L_{i},L_{j}]_{P}=\varepsilon _{ijk}L_{k}$. For the particle on the curved
surface, we have after calculations,%
\begin{equation}
\lbrack G_{i},G_{j}]_{D}=\varepsilon _{ijk}\left\{ G_{k}-x_{k}\tau \mathbf{%
x\cdot p+}\left( x_{k}\kappa -n_{k}\right) \mathbf{n\cdot G}\right\} .
\label{gg}
\end{equation}%
It reduces to the $[G_{i},G_{j}]_{D}=\varepsilon _{ijk}G_{k}$ for both
relations $\tau \mathbf{x\cdot p}=0$ and $\left( x_{k}\kappa -n_{k}\right) 
\mathbf{n\cdot G}=0$ come true. Clearly, for\ particle moves in the free
space, or in central force potential, or on the sphere, $\mathbf{G}$ is
identical to $\mathbf{L}$\textbf{.}

Next, we have following equations of motion for $\mathbf{x},\mathbf{p}$ and $%
\mathbf{G}$, respectively,%
\begin{align}
\frac{d\mathbf{x}}{dt}& \equiv \lbrack \mathbf{x},H]_{D}=\frac{\mathbf{p}}{%
\mu },  \label{xh} \\
\frac{d\mathbf{p}}{dt}& \equiv \lbrack \mathbf{p},H]_{D}=-\frac{\mathbf{n}}{%
\mu }(\mathbf{p\cdot \nabla n\cdot p}),  \label{ph} \\
\frac{d\mathbf{G}}{dt}& \equiv \lbrack \mathbf{G},H]_{D}=-\left( \mathbf{x}%
\times \mathbf{n}\right) \frac{\mathbf{p\cdot \nabla n\cdot p}}{\mu }\equiv 
\mathbf{T.}  \label{lh}
\end{align}%
A important property of vector $\mathbf{T=}d\mathbf{G/}dt$ is that it lies
on the tangential surface, for we have $\mathbf{n}\cdot \mathbf{T}=0$.
Relations (\ref{xh})-(\ref{lh}) are revealing but somewhat trivial. In
contrast, the consequence of these relations is significant in quantum
mechanics, as we see shortly.

\subsection{Quantum conditions of the constrained system}

The scheme of the canonical quantization hypothesizes that\ in general the
definition of a quantum commutator for any variables $f$ and $g$ is given by 
\cite{dirac2}, 
\begin{equation}
\lbrack f,g]=i\hbar O\left\{ [f,g]_{D}\right\}   \label{quantization}
\end{equation}%
in which $O\left\{ f\right\} $ stands for the quantum operator corresponding
to the classical quantity $f$. The \textit{fundamental quantum conditions}
are $[x_{i},x_{j}],[x_{i},p_{j}]$ and $[p_{i},p_{j}]$. For a particle moves
in the free space, we have two fundamental operators in quantum mechanics,
which in the configuration representation are position $\mathbf{x}$ and
momentum $\mathbf{p}=-i\hbar \nabla $. For a particle moves on a surface,
there is in general a great difficulty in getting the momentum operator \cite%
{ogawa1,ogawa2,ikegami}, because we run into the notorious operator-ordering
difficulty of momentum and position operators in $O\left\{
(n_{j}n_{i,k}-n_{i}n_{j,k})p_{k}\right\} \left( =[p_{i},p_{j}]/(i\hbar
)\right) $ from (\ref{pp}). Even worse is not the ambiguities in defining
the Hamiltonian operator, but the Schr\"{o}dinger-Podolsky-DeWitt approach
is not able to give the correct form of the Hamiltonian operator no matter
what form of the momentum operator is obtained \cite{ikegami}. Thus the
commutators $[p_{i},p_{j}]$ must be excluded from the so-called fundamental
quantum conditions for they contain severe vagueness. Thus, we should search
for \textit{quantum conditions} beyond the usual fundamental ones. A
straightforward enlargement of the quantum conditions is to simply follow
the hypothesis given by (\ref{quantization}) to include all $[f,H]$ as $f=%
\mathbf{x}$, $\mathbf{p}$ and $\mathbf{G}$ to simultaneously determine the
operators $\mathbf{p}$ and $H$. It is fruitless at all, because there are
much depressing operator-ordering difficulties in $O\left\{ \mathbf{n}(%
\mathbf{p\cdot \nabla n\cdot p})\right\} /\mu \left( =-[\mathbf{p}%
,H]/(i\hbar )\right) $ from (\ref{lh}) and $O\left\{ \left( \mathbf{x}\times 
\mathbf{n}\right) \mathbf{p\cdot \nabla n\cdot p}\right\} /\mu \left( =-[%
\mathbf{G},H]/(i\hbar )\right) $. To surmount these difficulties, we note
following vanishing relations,%
\begin{equation}
\mathbf{n}\cdot \lbrack \mathbf{x},H]_{D}=\mathbf{n}\cdot \frac{\mathbf{p}}{%
\mu }=0,\mathbf{n}\cdot \lbrack \mathbf{G},H]_{D}=\mathbf{n}\cdot \mathbf{F}%
=0\text{ and }\mathbf{n}\times \lbrack \mathbf{p},H]_{D}=0.  \label{zero}
\end{equation}%
The resultant quantum conditions\textit{\ free from the operator-ordering
difficulty} are given by,%
\begin{align}
\lbrack x_{i},x_{j}]& =0,\text{ }  \label{xxq} \\
\lbrack x_{i},p_{j}]& =i\hbar \left( \delta _{ij}-n_{i}n_{j}\right) ,
\label{xpq} \\
\lbrack \mathbf{x},H]& =i\hbar \frac{\mathbf{p}}{\mu },  \label{xhq} \\
\mathbf{n}\cdot \lbrack \mathbf{x},H]+[\mathbf{x},H]\cdot \mathbf{n}& =\frac{%
i\hbar }{\mu }(\mathbf{n\cdot p+p\cdot n})=0,  \label{zero1} \\
\mathbf{n}\times \lbrack \mathbf{p},H]+[\mathbf{p},H]\times \mathbf{n}& =0,
\label{zero2} \\
\mathbf{n}\cdot \lbrack \mathbf{G},H]+[\mathbf{G},H]\cdot \mathbf{n}&
=i\hbar \left( \mathbf{n\cdot F+F\cdot n}\right) =0.  \label{zero3}
\end{align}%
\textit{\ } The Hamiltonian operator must take the following form for it in
classical limit reduces to the classical Hamiltonian $H=p^{2}/2\mu $, 
\begin{equation}
H=-\frac{\hbar ^{2}}{2\mu }\nabla _{LB}^{2}+\mathbf{\alpha }(x)\cdot \nabla
_{s}+V_{G},  \label{trialh}
\end{equation}%
where $\mathbf{\alpha }(x)$ and $V_{G}$ are functions that go over to zero
not only in classical limit but also for free motion in flat space, which in
general does not have an analog in classical mechanics, and $\nabla
_{LB}^{2}=\nabla _{s}\cdot \nabla _{s}$ is the Laplace-Beltrami operator
which is the dot product of the gradient operator $\nabla _{s}$ on the
surface.

The first condition (\ref{xxq}) sets the configuration representation with
Cartesian coordinates. The second condition (\ref{xpq}) gives the essential
part of the momentum $\mathbf{p}$ is the gradient $\nabla _{s}$ on the
surface, 
\begin{equation}
\mathbf{p}=-i\hbar \left( \nabla _{s}+\mathbf{\beta }(x)\right) ,
\label{genp}
\end{equation}%
where $\mathbf{\beta }(x)$ is an undetermined vector function. Substituting (%
\ref{genp}) and (\ref{trialh}) into the third condition (\ref{xhq}), the
momentum operator $\mathbf{p}$ and Hamiltonian operator $H$ becomes,
respectively \cite{liu13-1,liu11}, 
\begin{equation}
\mathbf{p}=-i\hbar \left( \nabla _{s}+\frac{M}{2}\mathbf{n}\right) 
\label{gm}
\end{equation}%
where $M$ is a sum of two principal curvatures $R_{1}^{-1}$ and $R_{2}^{-1}$%
, ($R_{1}^{-1}+R_{2}^{-1}$) the mean curvature at point $\mathbf{x}$ on the
surface $S^{2}$ 
\begin{equation}
H=\frac{p^{2}}{2\mu }+V_{G}-\frac{\hbar ^{2}M^{2}}{8\mu },  \label{trialh1}
\end{equation}%
where $\mathbf{\alpha }(x)=0$ in Eq. (\ref{trialh}). It is easily to verify
that the fourth and fifth conditions (\ref{zero1}) and (\ref{zero2}) are
automatically satisfied whatever form\ of potential $V_{G}$ is. Lastly, let
us calculate the $\mathbf{n}\cdot \lbrack \mathbf{G},H]+[\mathbf{G},H]\cdot 
\mathbf{n}$, and after a lengthy but straightforward manipulation, we arrive
at,%
\begin{equation}
V_{G}=\frac{\hbar ^{2}M^{2}}{8\mu }-\frac{\hbar ^{2}}{4\mu }%
(M^{2}-2K)+\varphi =-\frac{\hbar ^{2}}{2\mu }\left\{ \left( \frac{M}{2}%
\right) ^{2}-K\right\} +\varphi ,  \label{geomp}
\end{equation}%
where $K$ is the gaussian curvature which is the product of the two
principal curvatures as $\left( R_{1}R_{2}\right) ^{-1}$, and function $%
\varphi $ satisfies following differential equation,%
\begin{equation}
\mathbf{n}\times \nabla \varphi =0.  \label{phi}
\end{equation}%
It means that $\nabla \varphi $ is parallel to the normal direction $\mathbf{%
n}\equiv \nabla f(x)$, and we have $\nabla \varphi =\Phi (x)\nabla f(x)$
with $\Phi (x)$ being the magnitude of gradient of function $\varphi $.
Since $\left\vert \nabla f(x)\right\vert =1$, we have thus $\Phi (x)=$ $\pm
\left\vert \nabla \varphi \right\vert $. So, the function $\varphi $ defines
a surface whose normal vector is identical to the surface $f(x)=0$. So, the
new surface is identical to $f(x)=0$, but takes another form $\varphi \left[
f(x)\right] =0$, i.e., we have $\varphi =0$. The quantum Hamiltonian
operator turns out to be,%
\begin{equation}
H=\frac{p^{2}}{2\mu }-\frac{\hbar ^{2}}{2\mu }\left( M^{2}-2K\right) =-\frac{%
\hbar ^{2}}{2\mu }\left( \nabla _{LB}^{2}+\left( \frac{M}{2}\right)
^{2}-K\right) .
\end{equation}%
In the first expression, the curvature-induced potential is negative
definite for we have $-\hbar ^{2}/(2\mu )\left( R_{1}^{-2}+R_{2}^{-2}\right) 
$, whereas in the second expression, the curvature-induced potential is
semi-negative definite for we have $-\hbar ^{2}/(8\mu )\left(
R_{1}^{-1}-R_{2}^{-1}\right) ^{2}$, which is the so-called geometric
potential,%
\begin{equation}
V_{G}=-\frac{\hbar ^{2}}{2\mu }\left\{ \left( \frac{M}{2}\right)
^{2}-K\right\} .
\end{equation}

\subsection{Further comments on the commutators $[p_{i},p_{j}]$}

The naive utilization of the relation (\ref{pp}), i.e., $O\left\{
(n_{j}n_{i,k}-n_{i}n_{j,k})p_{k}\right\} =[p_{i},p_{j}]/(i\hbar )$, is
highly controversial topic to construct momentum. For instance, once can
assume $O\left\{ (n_{j}n_{i,k}-n_{i}n_{j,k})p_{k}\right\} \equiv
c_{1}n_{j}n_{i,k}p_{k}+c_{2}p_{k}n_{j}n_{i,k}+c_{3}n_{j}p_{k}n_{i,k}+c_{4}n_{i,k}p_{k}n_{j}-(i\leftrightarrow j)
$ where $c_{1}+c_{2}+c_{3}+c_{4}=0$ \cite{ogawa1}. But, because $n_{j}$ and $%
n_{i,k}$ contain various functions of coordinates of $x$, $y$ and $z$, e.g.,
we can have $n_{j}=x_{j}\left( x^{2}+y^{2}+z^{2}\right) ^{-1/2}$, then
operators $p_{k}n_{j}$ in the $4$th term $n_{i,k}p_{k}n_{j}$ in above
decomposition of $O\left\{ (n_{j}n_{i,k}-n_{i}n_{j,k})p_{k}\right\} $ can at
least be further decomposed as $p_{k}n_{j}=d_{1}\left(
x^{2}+y^{2}+z^{2}\right) ^{-1/2}p_{k}x_{j}+d_{2}x_{j}p_{k}\left(
x^{2}+y^{2}+z^{2}\right) ^{-1/2}$ where $d_{1}+d_{2}=1$. No principle from
either physics or mathematics can be used to terminate this procedure. So,
the commutators $[p_{i},p_{j}]$ can hardly be members of the fundamental
quantum conditions. However, our geometric momentum turns out to satisfy the
following relation, 
\begin{equation}
O\left\{ (n_{j}n_{i,k}-n_{i}n_{j,k})p_{k}\right\} =\frac{1}{2}\left(
(n_{j}n_{i,k}-n_{i}n_{j,k})p_{k}+p_{k}(n_{j}n_{i,k}-n_{i}n_{j,k})\right) .
\end{equation}%
Instead, our proposal is to reverse the quantization conditions $%
[p_{i},p_{j}]/(i\hbar )=O\left\{ (n_{j}n_{i,k}-n_{i}n_{j,k})p_{k}\right\} $
and to construct a function containing momentum operators such that we have, 
\begin{equation}
\mathbf{n}\cdot \mathbf{P+P}\cdot \mathbf{n}=0,  \label{last}
\end{equation}%
where $P_{j}\equiv \mathbf{n}\cdot \lbrack \mathbf{p},p_{j}]+[\mathbf{p}%
,p_{j}]\cdot \mathbf{n}$. Since this relation (\ref{last}) alone is not yet
able to give the momentum operators, we see that the commutators $%
[p_{i},p_{j}]$ are not fundamental quantum conditions either.

\section{Conclusions and discussions}

The quantum conditions given by the straightforward applications of the
equation (\ref{quantization}) are not always fruitful, even misleading. For
the particle on the curved surface, in order to obtain the geometric
potential predicted by the so-called thin-lay quantization, a proper
enlargement of the quantum conditions turns out to be compulsory to contain
positions, momenta, orbital angular momentum and Hamiltonian. What is more,
a construction of unambiguous quantum conditions out of the equation (\ref%
{quantization}) proves inevitable. Combining the\ enlargement and the
construction, we obtain the geometric potential. Since momentum in the
orbital angular momentum is the geometric one, which only in some special
cases reduces to the usual one, we can call the orbital angular momentum the
geometric angular momentum. Even the present paper deals with only the
two-dimensional curved surface, we conjecture that our method can be used
for particle on an arbitrarily dimensional curved surface, which is still
under investigation.

Finally, we would like to point out that there are other forms of the
enlargement and the construction in literature, for instance Refs. \cite%
{bender,Deriglazov}, but they were devised to serve entirely different
purposes.

\begin{acknowledgments}
This work is financially supported by National Natural Science Foundation of
China under Grant No. 11675051.
\end{acknowledgments}

\end{document}